\documentclass[pra,twocolumn,showpacs,superscriptaddress,amsmath,amssymb]{revtex4-1}

\usepackage{graphicx}

\begin{document}

\title{Simulating strongly-correlated multiparticle systems in a truncated Hilbert space}

\author{Thomas Ernst}
\affiliation{%
New Zealand Institute for Advanced Study and Centre for Theoretical Chemistry and Physics, Massey University, Private Bag 102904, North Shore, Auckland 0745, New Zealand
}%
\author{David W. Hallwood}
\affiliation{%
New Zealand Institute for Advanced Study and Centre for Theoretical Chemistry and Physics, Massey University, Private Bag 102904, North Shore, Auckland 0745, New Zealand
}
\author{Jake Gulliksen}
\affiliation{%
New Zealand Institute for Advanced Study and Centre for Theoretical Chemistry and Physics, Massey University, Private Bag 102904, North Shore, Auckland 0745, New Zealand
}
\author{Hans-Dieter Meyer}
\affiliation{Theoretische Chemie, Physikalisch-Chemisches Institut, 
Universit\"at Heidelberg, Im Neuenheimer Feld 229, D-69120 Heidelberg, Germany}

\author{Joachim Brand}
\affiliation{%
New Zealand Institute for Advanced Study and Centre for Theoretical Chemistry and Physics, Massey University, Private Bag 102904, North Shore, Auckland 0745, New Zealand
}%

\date{\today}

\begin{abstract}
Representing a strongly interacting multiparticle wave function in a finite product basis leads to errors. 
Simple rescaling of the contact interaction can preserve the low-lying energy spectrum and long-wavelength structure of wave functions in one-dimensional systems and thus correct for the basis-set truncation error. The analytic form of the rescaling is found for a two-particle system where the rescaling is exact. A detailed comparison between finite Hilbert space calculations and exact results for up to five particles show that rescaling can significantly improve the accuracy of numerical calculations in various external potentials. In addition to ground-state energies, the low-lying excitation spectrum, density profile and correlation functions are studied. The results give a promising outlook for numerical simulations of trapped ultracold atoms.
\end{abstract}

\maketitle

\section{\label{intro}Introduction}
Numerically solving the many-particle Schr\"odinger equation in one spatial dimension with short-range interaction requires a truncated Hilbert space to make the calculations computationally tractable. This introduces inaccuracies especially in strongly correlated regimes, systems with many degeneracies, and near avoided crossings. Here, we demonstrate a simple method that significantly improves results by rescaling the interaction strength.

Understanding the complexities of many-body quantum physics remains a grand challenge, in particular with regard to excited states and dynamical problems. Ultracold atoms have recently become a testbed for both analytic and computational approaches to understanding quantum many-body physics. These systems can be prepared with few to millions of particles and the experimental control over interactions, system size, and quantum states is rapidly improving \cite{greiner2002quantum,RevModPhys.82.1225,grünzweig2010near}.

However, exact theoretical results are rare, especially for few-particle systems. Approximate approaches, like the mean-field Gross-Pitaevskii equation, accurately model a large number of weakly interacting atoms. Now, there is a considerable amount of research that focuses on the creation of entanglement in ultracold atomic systems. These require calculations at a single-particle level. Much success with fixed basis-set expansions has been achieved using Wannier functions of an optical lattice leading to the Bose-Hubbard model \cite{PhysRevLett.81.3108} and with harmonic oscillator eigenfunctions \cite{PhysRevA.75.013614} or a plane-wave basis \cite{Kanamoto,PhysRevA.82.063623} for trapped systems. These approaches  are limited by accessing only a finite Hilbert space. Methods like time-evolving block decimation \cite{PhysRevLett.91.147902} and time-dependent density matrix renormalization \cite{PhysRevLett.93.076401,1742-5468-2004-04-P04005} exploit spatial entanglement properties to allow for time-dependent simulations of larger multiparticle problems.

A different approach is taken by the multiconfigurational time-dependent Hartree (MCTDH) method \cite{meyer1990multi,bec00:1,meyer2009multidimensional}. It variationally optimizes a set of single-particle functions, that define the accessible Hilbert space at any given moment in time. It has been successfully applied to ultracold gases \cite{PhysRevA.74.053612,PhysRevA.78.013621,PhysRevA.78.013629}. Even so, the truncated basis leads to errors in strongly  interacting systems. Here, we analyze the origin of errors that arise from basis-set truncation, suggest a scheme to compensate for them, and benchmark the performance of the scheme.

Trapped bosonic atoms confined to a single spatial dimension undergo a crossover from a Bose-Einstein condensate to the fermionized and strongly correlated Tonks-Girardeau (TG) gas \cite{girardeau1960} as
repulsive short-range interactions are tuned from weak to very strong. This system is well suited to benchmark computational many-body methods as exact results are available from
integrable models in the limits of infinite interactions (TG \cite{girardeau1960}) or vanishing external potential (Lieb-Liniger model \cite{lieb1963exact}). Previous
calculations of the ground state with exact diagonalization in a fixed product basis \cite{PhysRevA.75.013614,PhysRevA.78.013604,epjdHaoChen09}  and with the MCTDH method \cite{PhysRevA.74.053612,zoe06:063611} have shown a peculiar feature: Specific finite values of the interaction strength in these calculations provide a reasonable approximation of the completely fermionized state, which in exact theory is only reached for infinite interaction. Our calculations show that larger values of the interaction produce worse results in the approximate calculation for the energy and lead to spuriously enhanced density oscillations. 
In this paper we provide a clear explanation of this effect based on exact results for two particles. 

We find that the origin of the problem lies in the truncation of the Hilbert space. Previously, elaborate schemes have been developed to generate effective Hamiltonians that remove this problem, however these techniques are cumbersome and require careful application \cite{suzuki1980convergent,PhysRevA.76.063613,PhysRevLett.100.230401,PhysRevA.79.012707}.  A simpler, yet powerful method is desirable to accurately describe different interacting regimes for a variety of applications \cite{albiez2005direct,carr2010dynamical,brennen1999quantum,tretyakov2006cold}.
Here we present a rescaling method that increases accuracy or decreases the required Hilbert space by mapping the interaction in the simulations to the physical interaction. Therefore numerical simulations can be sped up and used to approach previously unaccessible regimes and systems. 

In this paper we initially present an exact rescaling procedure for two particles in a one-dimensional periodic box as described previously in Ref.\ \cite{PhysRevA.82.063623}. A more general scheme is proposed that only requires the knowledge of the numerical value of the interaction strength at the TG energy. This approach is closely related to the known renormalization of the two-body $T$ matrix but extends this method into the strongly interacting regime. The approach is tested for different system geometries, particle numbers and external potential shapes. The results are compared to analytical results for two atoms in a harmonic potential \cite{busch1998}, the Lieb-Liniger description of atoms confined to a one-dimensional periodic system \cite{lieb1963exact} and infinitely strong interacting atoms in one dimension described by a TG gas \cite{girardeau1960}.

\section{\label{sec:level0}Theoretical background} 
\subsection{Finite basis-set expansion}
Consider $N$ interacting quantum particles in one dimension subject to an external potential $V_{\rm ext}(x,t)$, which may depend on time:
\begin{equation}\label{eq:H1}
H_{mb}=\sum_{i=1}^N h_{i}(x_i) + \sum_{i<j} g\delta(x_i-x_j) ,
\end{equation}
where $h_{i}=-\frac{\hbar^2}{2m}\frac{\partial^2}{\partial x_i^2} + V_{\rm ext}(x_i,t)$ is the one-body Hamiltonian and for simplicity we assume that all particles have the same mass $m$ and interact by contact interactions of strength $g$. This model describes, e.g., quantum gases of ultracold atoms in an elongated and tightly confining trapping potential \cite{Olshanii}. Although we specifically deal with identical bosons later on, at this point considerations are not restricted to a specific quantum statistics and the Hamiltonian (\ref{eq:H1}) could describe single- or multicomponent Bose or Fermi gases or mixtures thereof. All these possibilities are of interest and related to actual or possible experimental scenarios.

The true $N$-particle wave function can be expanded in the form
\begin{equation} \label{eq:fbsexp}
 \Psi(x_1,\ldots,x_N, t) =  \sum_{J} A_J(t) \Phi_J(x_1,\ldots,x_N) ,
\end{equation}
in a basis consisting of products of single-particle wave functions
\begin{equation}\label{eq:product}
\Phi_J(x_1,\ldots,x_N) = \prod_{k=1}^{N} \phi_{j_k}(x_k) ,
\end{equation}
where we have introduced the multi-index $J=(j_1,\ldots,j_N)$ for the set of $N$ single-particle indices. The single-particle wave functions (or {\em mode} functions) are mutually orthonormal by $\int \phi^*_{j}(x)\phi_k(x)dx=\delta_{jk}$. 

In practice we chose a finite set of $M$ single-particle functions to define the finite basis-set expansion. The size of the finite basis of multiparticle Hilbert space is nominally $M^N$, although for identical particles this number can be significantly reduced by accounting for bosonic or fermionic exchange symmetry in a well-known manner by symmetrizing the products to permanents or Slater determinants, respectively.

\begin{figure}
\includegraphics[width=1.0\columnwidth,angle=-0]{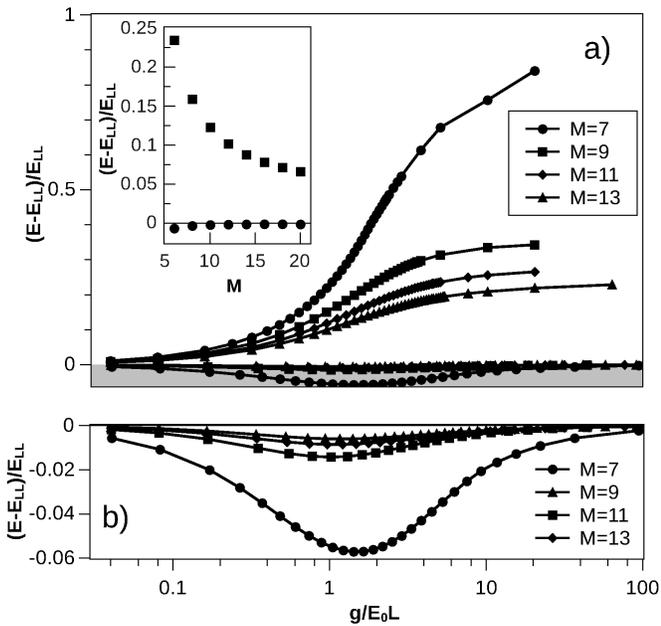}
\caption{\label{fig:before_after_varM5p} Relative deviation of the approximate ground-state energy from  the exact Lieb-Liniger (LL) result \cite{lieb1963exact} before rescaling (white area of panel a) and after rescaling (gray area of panels a and b) for five particles in a box of length $L$ with periodic boundary conditions. Rescaling reduces the maximum errors from 23\% to 0.6\% ($M=13$). The inset shows the relative energy deviation vs the number of modes $M$ for three particles at $g=1.9099$. Results for the truncated Hamiltonian without rescaling (squares) converge very slowly while rescaling (circles) significantly reduces the errors already for small $M$.}
\end{figure}

The standard procedure for approximating the full problem (\ref{eq:H1}) is to simply truncate the Hilbert space to span a finite basis and represent the Hamiltonian as a finite matrix with elements
\begin{equation}
H_{IJ} =  \langle \Phi_I|H_{mb}|\Phi_J\rangle .
\end{equation}
We denote the solution of the corresponding matrix Schr{\"o}dinger equation for the coefficient vector with $\bar{A}_{J}$ and the eigenvalues of the truncated matrix $\bar{E}_\nu$. These approximate the exact coefficient vector ${A}_{J}$ and eigenvalues $E_\nu$, which form part of the spectrum of  $H_{mb}$. Since the basis-set truncation can be understood as a variational procedure with a restricted variational space, the approximate ground-state energy $\bar{E}_G \ge E_G$ is an upper bound for the true one. By increasing the number of modes $M$ and thereby computational space, the approximation improves and the approximate energy converges to the correct one. However, for an interacting system this convergence is painfully slow, as seen in the inset in Fig.~\ref{fig:before_after_varM5p}.

The purpose of this paper is to improve the approximations for a given, finite, basis-set expansion. While previous works have explored elaborate procedures to approximate an effective finite Hamiltonian with the exact eigenvalues $E_\nu$ \cite{suzuki1980convergent,PhysRevA.76.063613,PhysRevLett.100.230401}, we show that it is possible to obtain significant improvements by simply rescaling the interaction parameter $g$. 

\subsection{Two particles on a ring}\label{sec:twoparticles}
For two particles without external potential an  analytic expression for the rescaled interaction constant $\tilde{g}$ can be found that reproduces the correct ground-state energy and the Fourier components of the wave function up to a cutoff. The rescaling procedure is thus exact for this case.

\begin{figure}
\includegraphics[width=1.0\columnwidth,angle=-0]{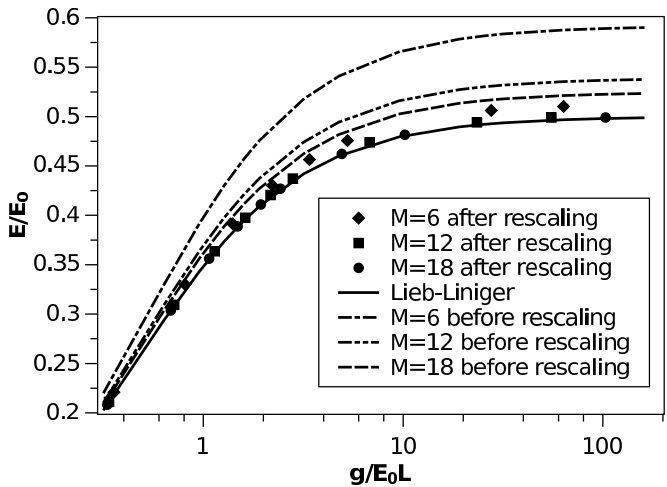}
\caption{\label{fig:2p-rescaling}Ground-state energies for two interacting bosons in a periodic box before (lines) and after (symbols) rescaling according to Eqs.~(\ref{eq:rescaling}) and (\ref{eq:rescanal}) compared with the exact (Lieb-Liniger) 
energy (solid line). The calculated energies have been obtained via exact diagonalization of the truncated Hamiltonian matrix with different numbers of modes $M$. Clearly, rescaling of the interaction reduces the errors in the energies significantly.}
\end{figure}

Consider $N=2$ bosons in a one-dimensional box with length $L$ and periodic boundary conditions. This model might be realized with ultracold atoms in a tightly focused ring trap. The Hamiltonian (\ref{eq:H1}) simplifies to 
\begin{equation}
H_{2p}=-\frac{\hbar^2}{2m}\left( \frac{\partial^2}{\partial x_1^2} +\frac{\partial^2}{\partial x_2^2} \right)+ g\delta(x_1-x_2)\label{eq:2pHamiltonian}
\end{equation}
where $x_i$ is the spatial coordinate of atom $i$ and $g$ the physical interaction strength. A finite basis-set expansion for this system is
\begin{equation}
\psi(x_1,x_2) = \sum_{k_1,k_2=-(M-1)/2}^{(M-1)/2} C_{k_1,k_2} e^{i\frac{2\pi}{L}k_1x_1} e^{i\frac{2\pi}{L}k_2x_2} ,\label{eq:2pWaveFunction}
\end{equation}
where we assume the number of momentum modes $M$ in the expansion to be odd, for simplicity.
First consider the case for the full Hilbert space with $M=\infty$. To find the ground state we substitute Eq.~(\ref{eq:2pWaveFunction}) and Eq.~(\ref{eq:2pHamiltonian}) into the Schr{\"o}dinger equation $(E-H_{2p})\psi = 0$. 
Projecting to zero-momentum solutions by multiplying with $\exp[-i 2 \pi k(x_1-x_2)/L]$ and integrating over the particle coordinates leads to 
\begin{equation}
(E-2E_0 k^2)C_{k,-k}=\frac{g}{L}\sum^\infty_{q=-\infty} C_{q,-q}.
\label{eq:2pse}
\end{equation}
Here, $E_0 = 2\pi^2 \hbar^2 /(mL^2)$ is the smallest nonzero single-particle energy. The right-hand side of Eq.~(\ref{eq:2pse}) is independent of $k$ and therefore $C_{-q,q} \propto (E-2E_0q^2)^{-1}$ and
\begin{equation}
\frac{E_0 L}{g}=\sum^\infty_{q=-\infty} \frac{1}{(E/E_0-2q^2)}, \label{eq:g}
\end{equation}
which relates the exact energy $E$ to the physical interaction strength $g$. Employing the finite basis-set expansion limits the sum over $q$ to only $M$ terms and thus yields a different, approximate, value $\bar{E}$ for the energy. Here, we choose a different path and introduce the rescaled interaction strength $\tilde{g}$ by
\begin{align}
\frac{E_0 L}{\tilde{g}}&=\sum_{q=-(M-1)/2}^{(M-1)/2} \frac{1}{(E/E_0-2q^2)}, \label{eq:gtilde}\\
&=\frac{E_0 L}{{g}}+\frac{E_0 L}{g_0}, \label{eq:2psplit}
\end{align}
where the last equation defines the constant $g_0$. While Eq.~(\ref{eq:gtilde}) guarantees that $\tilde{g}$ yields the exact energy $E$ it is Eq.~(\ref{eq:2psplit}) that provides the rescaled interaction constant  $\tilde{g}$  as a function of $g$:
\begin{equation}
\tilde{g}=\frac{g}{1+g/g_0}.\label{eq:rescaling}
\end{equation}
Therefore, solving the Schr{\"o}dinger equation with this rescaled value $\tilde{g}$ instead of $g$ in the Hamiltonian   (\ref{eq:2pHamiltonian}) in the finite basis-set expansion gives the exact energy $E$ for the physical interaction strength $g$. Also the finite number of expansion coefficients $C_{k,-k}$ is identical to the full Hilbert space expansion and, according to Eq.~(\ref{eq:2pse}), is given by
\begin{equation}
C_{k.-k} = \frac{A}{E-2E_0 k^2} ,
\end{equation}
where 
\begin{equation}
A = \frac{g}{L}\sum^\infty_{q=-\infty} C_{q,-q} = \frac{\tilde{g}}{L}\sum_{q=-(M-1)/2}^{(M-1)/2} C_{q,-q}.
\end{equation}
The value of $g_0$ is found by expanding the sums in powers of $1/M$ from
\begin{equation} \label{eq:g0}
\frac{E_0 L}{g_0}=\frac{2}{M}+\frac{2E-E_0}{2E_0 M^3}+O(M^{-5}).
\end{equation}
For a large number of modes and sufficiently low energy $E$ [with the approximate condition $M^2\gg E/(2E_0)$] we thus find to leading order that $g_0$ is independent of the energy and given by
\begin{equation} \label{eq:rescanal}
 g_0 \approx \frac{1}{2}  M E_0 L.
\end{equation}
Figure~\ref{fig:2p-rescaling} shows how the rescaling in approximation (\ref{eq:rescanal}) significantly improves the ground-state energy. For $M=6$ the energy after rescaling is only $3\%$ off the exact energy 
whereas before rescaling the difference was $\sim20\%$. 

The constant $g_0$ is readily interpreted from Eq.~(\ref{eq:rescaling}) as the value of the rescaled interaction $\tilde{g}$ where the physical interaction $g$ reaches infinity. Physically, this is the TG limit where the contact interactions are arbitrarily strong and the system is strongly correlated.  Our results thus indicate that a finite basis-set expansion with a finite value of the rescaled interaction constant $\tilde{g}=g_0$ can produce exact results for infinite values of the physical interaction strength. Since the energies and wave functions in the TG limit can also be calculated by the Bose-Fermi mapping \cite{girardeau1960}, this limit is an important reference point.

\subsection{Empirical rescaling} \label{sec:rescaling}

The analytical results of the preceding section motivate us to extend the idea of rescaling to more than two particles and to problems with arbitrary external potential, which, generally, cannot be solved analytically. We propose to replace the Hamiltonian (\ref{eq:H1}) by the rescaled version
\begin{equation}\label{eq:H1resc}
\tilde{H}_{mb}=\sum_{i=1}^N h_{i}(x_i) + \sum_{i<j} \tilde{g}\delta(x_i-x_j) ,
\end{equation}
where we have only changed the value of the interaction strength from the physical value $g$. Within a finite basis-set expansion we obtain a matrix with elements $\tilde{H}_{IJ}$, which is to be used for numerical simulation. The smallest eigenvalue is the approximate ground-state energy $\bar{E}_G(\tilde{g})$. For the rescaled interaction $\tilde{g}$ we continue to use Eq.~(\ref{eq:rescaling}) and determine the value of $g_0$ by requiring that the approximate ground-state energy at $g_0$ equals the exact energy of the TG limit:
\begin{equation} \label{eq:E0}
\bar{E}_G(g_0) = E_{\rm TG}.
\end{equation} 
A finite solution for $g_0$ can always be found since the approximate value of $\bar{E}_G(g)$  overestimates the real value and $\bar{E}_G(\tilde{g})$ is a monotonously growing function of $\tilde{g}$ with $\bar{E}_G(0)\le E_{\rm TG}$. The value of $E_{\rm TG}$ is equal to the ground-state energy of a system of noninteracting fermions due to the Bose-Fermi mapping theorem \cite{girardeau1960}. It is given by the sum of single-particle energies, which can be determined analytically or numerically to very high precision.

This rescaling procedure is very simple to implement, since only knowledge  of the value of the single constant $g_0$ is needed. We assume here for simplicity that $g_0$ is independent of the energy of the wave function and can be used for any set of (low-lying) excited states as well as time-dependent processes involving such states. As an approximation, this is consistent with the analytic findings for two particles from the previous section.

\subsection{Connection to $T$ matrix renormalization}

The rescaling approach introduced in Sec.~\ref{sec:rescaling} is closely connected to the well-known renormalization of the scattering $T$ matrix (for a readable account see Ref.~\cite{castin-2004-116}).
In this approach, the coupling constant of an effective many-body theory is renormalized to produce the correct $T$ matrix for the two-body scattering problem, which is known exactly.  
Indeed, Eq.~(\ref{eq:rescaling}) for the rescaled interaction strength together with  Eq. (\ref{eq:rescanal}) for the value of $g_0$ are identical to the result of Ref.~\cite{castin-2004-116} for the renormalized $T$ matrix of a high-momentum cutoff introduced there due to discretization of space (identifying the computational volume of momentum space $2\pi M/L$ with the Brillouin zone of Ref.~\cite{castin-2004-116}). This result was rigorously derived for weak interactions and high cutoffs. In Sec.~\ref{sec:twoparticles} we have extended the approach to the strongly interacting regime for two particles where now the rescaled interaction constant in principle becomes energy dependent. The extension to nonperturbative, strongly interacting multiparticle systems in Sec.~\ref{sec:rescaling} with Eq.~(\ref{eq:g}) for the value of $g_0$ is heuristic in nature. The purpose of the numerical studies reported in the following section is to establish the usefulness and to quantify the accuracy of this approach.

Rescaling improves numerical calculations by either increased accuracy or decreased computational effort for a given desired accuracy (CPU time $\sim M^3$ for our five particle simulations).
For $\tilde{g}> g_0$ we find unphysical results. Therefore, CPU times are further reduced as only results for $\tilde{g}\leq g_0$ are used in the rescaling.

Before detailing our numerical results and benchmarking, we introduce the MCTDH method suitable for general trapping potentials.  

\subsection{Variationally optimized basis}

The finite basis-set expansion (\ref{eq:fbsexp})  and (\ref{eq:product}) expands the multiparticle wave function in products of a given orthonormal set of single-particle functions. Optimizing this set variationally is the aim of the MCTDH method. This is achieved with a time-dependent variational ansatz with a fixed number of basis functions without further approximation \cite{meyer1990multi}, in contrast to the density matrix renormalization group approach, which further decomposes the problem into bipartite systems \cite{PhysRevLett.93.076401,1742-5468-2004-04-P04005} before finding an optimized basis.

The single-particle wave functions $\phi_j(x,t)$ are now allowed to be time dependent in addition to the time dependence of the coefficients  $A_J(t)$. The variational equations of motion for the coefficients and the single-particle wave functions are discretized in time (and space) and solved numerically with the MCTDH program package \cite{bec00:1,meyer2009multidimensional}.

Eigenenergies and eigenstates can be obtained by the relaxation method \cite{kos86:223},
i.~e., by propagation in negative imaginary time. Here we use a modification,
\emph{improved relaxation} \cite{mey06:179}, which can also be used in block form \cite{dor08:224109}
to simultaneously compute a set of eigenvectors. 
For a fixed basis of single-particle functions, the MCTDH method reduces to the exact diagonalization method. Therefore, the key advantage is that the MCTDH method always stays in a variationally  optimized basis. This generally ensures higher accuracy or keeps the Hilbert space as small as possible for a given accuracy. The MCTDH method is therefore effective at describing different trapping potentials. However, it requires a numerically unattainable number of single-particle functions to describe strongly interacting systems.

In practice, the single-particle wave functions are represented on a large number of primitive basis functions, the choice of which depends on the spatial geometry (typically 128 and 162 grid points in the harmonic oscillator discrete variable representation for the harmonic and the double-well potentials respectively and 155 points in the exponential discrete variable representation as per Ref.~\cite{bec00:1}). The setup of the MCTDH method for ultracold atom calculations follows Ref.~\cite{PhysRevA.74.053612}. In particular, the contact potential is replaced by a narrow Gaussian for practical purposes. We have carefully studied the dependence of our calculations on the parameters of the primitive basis and the Gaussian interaction to make sure that the results presented here are converged and the influence of these further approximations is negligible. 

\section{\label{ring}Bosons in a ring}

We first benchmark the rescaling method of Sec.~II.C. for the Lieb-Liniger model \cite{lieb1963exact} of bosons in a box with periodic boundary conditions, where exact eigenstate energies are easily available for any values of the interaction strength and particle number. Figure \ref{fig:before_after_varM5p} shows the relative deviation from the exact ground-state energy obtained from MCTDH calculations before and after rescaling. Rescaling is seen to significantly reduce the error. The rescaled results reproduce the exact energy in the TG limit by construction but the  improved accuracy for finite interaction strength is nontrivial. As expected, both the raw and rescaled data improve with increasing $M$. The maximal errors for, e.g., $M=13$ can be reduced from $\sim 23\%$ down to $\sim 0.6\%$, i.~e. a factor of 40. This is a significant improvement.

\begin{figure}
\includegraphics[width=1.0\columnwidth,angle=-0]{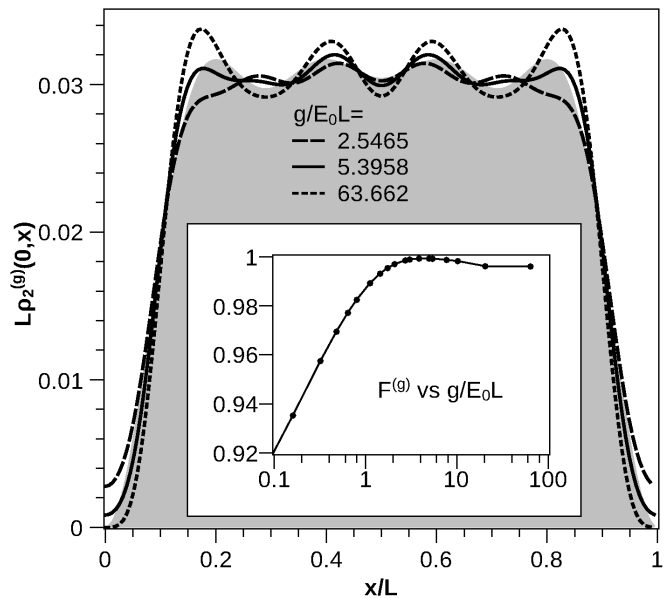}
\caption{\label{fig:density_difference_varg_M13} Density-density correlation function for five particles and $M=13$ single-particle functions in a periodic box. This plot compares  the reduced two-particle density $\rho_2^{(g)}$ of Eq.~(\ref{eq:2bd}) for different interaction strengths from MCTDH calculations without rescaling with the exact TG expression (the area below this curve is shaded). We find a good agreement with the exact solution close to $g_0=5.3958E_0 L$. This is shown in the inset where the overlap $F^{(g)}(\rho_2^{(g)},\rho_2^{TG})$ of Eq.~(\ref{eq:F}) is plotted as a function of the interaction $g$. There, the maximum occurs close to $g_0$.}
\end{figure}

In order to test how well the wave function is approximated in the TG limit we investigate the reduced pair density 
\begin{equation}\label{eq:2bd}
\rho_2^{(g)}(0,x)=\int\cdots\int dx_3\ldots dx_N |\Psi^{(g)}(0,x,x_3,\ldots ,x_N,t)|^2 .
\end{equation}
The results of MCTDH calculations are shown in Fig.~\ref{fig:density_difference_varg_M13} for different values of the interaction $g$ along with the exact solution in the TG limit $\rho_2^{TG}(0,x) = [(2 + \cos x) \sin^2 \frac{x}{2}] /3\pi^2$ \cite{girardeau1960}. Agreement between the finite basis-set expansion at the value of $g_0=5.3958E_0 L$ and the exact solution is reasonable. The discrepancies can well be explained by the absence of short-wavelength modes in the finite expansion. Most importantly, the agreement between the approximate and exact functions clearly worsens for interaction strengths larger than $g_0$, where the structure in the finite basis-set calculation overshoots.

In order to quantify the quality of the approximation, we define the  overlap with the exact solution at the TG regime by
\begin{align}\label{eq:F}
& F^{(g)}(\rho_2^{(g)},\rho_2^{TG})= \nonumber \\ 
& \frac{\int \rho_2^{TG}(0,x)\rho_2^{(g)}(0,x)dx}{\sqrt{\int \rho_2^{(g)}(0,x)\rho_2^{(g)}(0,x)dx} \sqrt{\int \rho_2^{TG}(0,x)\rho_2^{TG}(0,x)dx}} ,
\end{align}
which is shown in the inset of Fig.~\ref{fig:density_difference_varg_M13}. We find that the maximum of $F^{(g)}$ is reached near $g_0$, which justifies the choice of $g_0$ through Eq.~(\ref{eq:E0}).

\begin{figure}
\includegraphics[width=1.0\columnwidth,angle=-0]{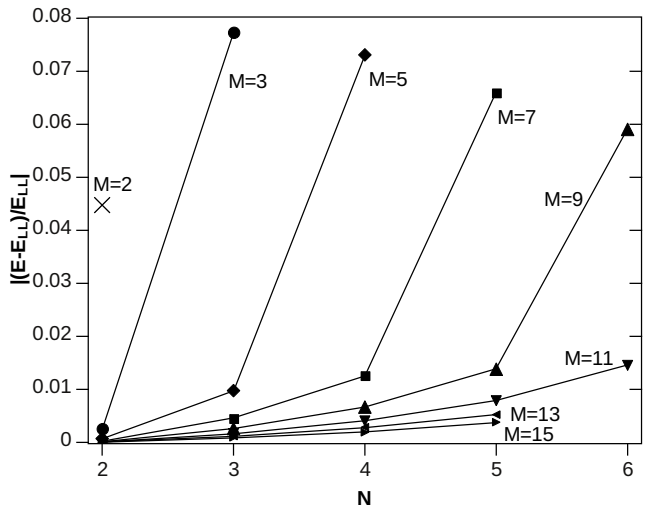}
\caption{\label{fig:scaling-particles} Maximal relative deviation of the energy after rescaling from the exact result $E_{LL}$ over the range of interaction strengths  $0<g<\infty$. The particle number and box length $L$ are varied while the density is kept fixed. 
}
\end{figure}

In order to study the scaling with particle number, we have performed exact diagonalization calculations for $N=2\ldots6$ particles in a ring of length $L=N/\tilde{n}$, where the particle density $\tilde{n}$ was fixed to the same value as in Fig.~\ref{fig:density_difference_varg_M13}. Figure \ref{fig:scaling-particles} shows the maximal relative deviation of the rescaled energy as a function of particle number.
The results show that retaining $M=2N-1$ single-particle functions gives a maximal relative error  bounded by $|E/E_{LL}-1|\lesssim 1.5\%$ for any value of $g>0$. 
Although for clarity only odd numbers of single-particle modes $M$ are shown in Fig~\ref{fig:scaling-particles}, we have verified that these findings also hold for even $M$.

If we assume that the scaling $M=2N-1$ is sufficient to maintain accuracy uniformly over the whole range of interaction strengths also for larger particle numbers, we still find that the numerical effort of MCTDH or exact diagonalization studies grows exponentially with $N$. This limits studies of this sort to fewer than 10 particles with current-day computers. 

\section{\label{HO}Bosons in a harmonic potential}

In order to test the rescaling method in the presence of an external potential we consider a harmonic trapping potential
\begin{equation}
 V_{\rm ext}(x)/E_{0,H}=\frac{1}{2}\left(\frac{x}{L_H}\right)^2.
\end{equation}
For this and the following section we use as natural energy and length scales the harmonic oscillator level spacing $E_{0,H}=\hbar\omega$ and  $L_H=\sqrt{\frac{\hbar}{m\omega}}$, respectively, where $\omega$ is the harmonic oscillator frequency.

\begin{figure}
\includegraphics[width=1.0\columnwidth,angle=-0]{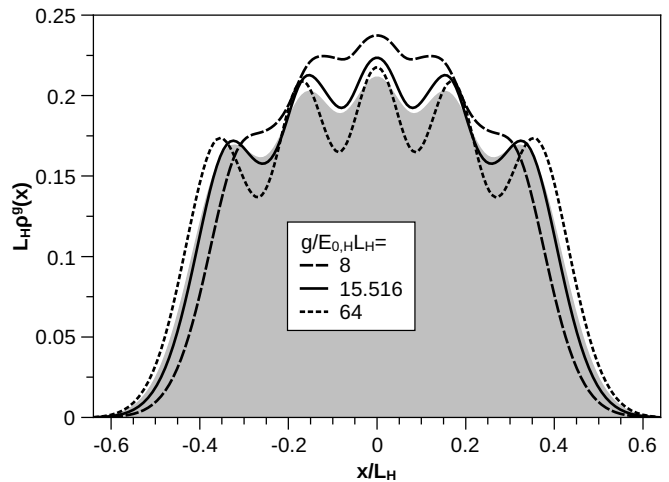}
\caption{\label{fig:HO-density} Density of five particles in a harmonic potential for different interaction strengths without rescaling and $M=13$. Analogous to Fig.~\ref{fig:density_difference_varg_M13} the best agreement with the exact TG result, which is given by the shaded areas, can be found at $g=g_0$, where $g_0 =15.516E_0 L$.}
\end{figure}

Figure~\ref{fig:HO-density} compares the single-particle density of the TG wave function with MCTDH simulations for five particles at different values of the interaction strength without rescaling. The agreement between the exact  and the finite basis-set results at the value of the interaction strength $g_0=15.516\ldots E_{0,H} L_H$ is much better than for larger values of $g$. This result is analogous to the ring case and supports our proposition that $g_0$ from Eq.~(\ref{eq:E0}) provides a good approximation for the wave function.

\begin{figure}
\includegraphics[width=1.0\columnwidth,angle=-0,clip]{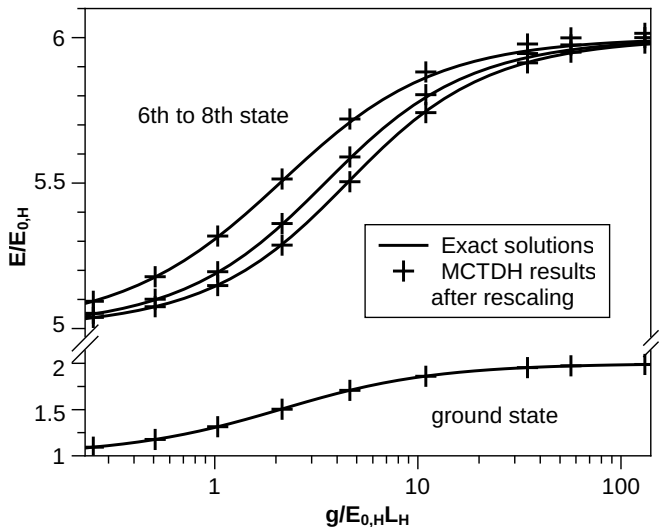}
\caption{\label{fig:HO-energies} Energies of the ground state and the sixth to eighth excited states with $M=15$ of two bosons in a harmonic well after rescaling compared to the exact results from Eq.~(\ref{eq:HOexact}). 
The results clearly show the high accuracy of the rescaled results. The largest error occurs for the seventh excited state where it is decreased from $\sim2\%$  before rescaling (not shown on this graph) to $\sim 0.3\%$ after rescaling.}
\end{figure}

Full analytical solutions are available for two particles in a harmonic potential. The exact energies for the ground and excited states are found from the implicit equation \cite{busch1998}
\begin{equation} \label{eq:HOexact}
\frac{g}{E_{0,H} L_H}=2\sqrt{2}\frac{\Gamma(-\frac{E}{2E_{0,H}}+\frac{3}{4})}{\Gamma(-\frac{E}{2E_{0,H}}+\frac{1}{4})}.
\end{equation}
where $\Gamma$ is the gamma function. Figure~\ref{fig:HO-energies} compares the ground-state and a few excited-state energies with rescaled MCTDH results. The highest few states and the ground state have been chosen as they are more likely to show the largest errors within the block-diagonalization scheme (see below). We emphasize that a single value of $g_0 =29.75$ obtained from Eq.~(\ref{eq:E0}) was used for rescaling. The procedure not only maintains the correct order of excited states but also quantitatively describes the excited-state energies remarkably well.

\section{\label{DW}Bosons in a double well}

The rescaling method is further tested for the description of excited states of the TG gas in a double-well potential. Exact solutions are not available for interacting multiparticle systems in the double well except for the TG gas, where their exact energies are simply found from sums of single-particle energies (which can be determined with great accuracy) due to the Bose-Fermi mapping of Ref.~\cite{girardeau1960}.

\begin{figure}
\includegraphics[width=1.0\columnwidth,angle=-0]{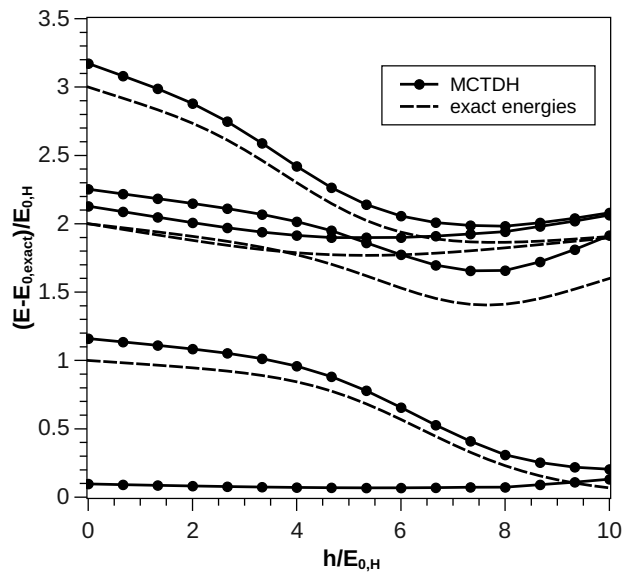}
\caption{\label{fig:double-well} Excited states of the Tonks-Girardeau (TG) gas in a double-well trap. Shown are the energies of the ground state and the first four excited states minus the exact TG ground-state energy for five bosons in a double-well potential vs barrier height $h/E_{0,H}$. The graph compares rescaled MCTDH calculations with $M=13$ single-particle functions (solid line with bullets) and exact TG energies (dashed line).
}
\end{figure}

We consider the double-well potential
\begin{equation}
V_{\rm ext}(x)/E_{0,H}=\frac{1}{2}\left(\frac{x}{L_H}\right)^2+\frac{h}{E_{0,H}}e^{-\frac{2}{3}(x/L_H)^2}
\end{equation}
with varying barrier height $h$ at the center.
Figure~\ref{fig:double-well} shows the energies for the ground and excited states from rescaled MCTDH calculations and analytical results. All MCTDH results were obtained in the TG regime at $\tilde{g} = g_0$. The value of $g_0$ has been determined for each value of $h$ from a dedicated ground-state calculation.

The results reported in Fig.~\ref{fig:double-well} were calculated using the block-improved-relaxation method \cite{dor08:224109} to efficiently determine several states simultaneously. 
Since an identical set of single-particle functions is used for the finite basis-set expansion for all states, the result is not as well variationally optimized for each individual state as the single-state calculation that was used to determine $g_0$. This explains the discrepancy of the numerical and exact ground-state energies seen in Fig.~\ref{fig:double-well}, which also provides a convenient order-of-magnitude estimate for the error of the block-improved-relaxation scheme. A shift to high energies for the whole set of numerical values compared to the exact ones of the same approximate magnitude is clearly observed in the data.

In addition, the second and third excited states in Fig.~\ref{fig:double-well}  should be degenerate  at $h=0$, which is not reproduced by the rescaled MCTDH results. By the Bose-Fermi mapping, these two states are identified as configurations $(1,1,1,1,0,0,1,0,\ldots)$ and $(1,1,1,0,1,1,0,0,\ldots)$ in the occupation number basis of fermions in the harmonic oscillator ($h=0$) potential. The two states involve different modes in the fermionic description and thus have different sensitivity to the finite set of available single-particle functions in the block-improved-relaxation method. This is a problem that the proposed rescaling method cannot fix.

The problem vanishes when more single-particle functions are used, but this is costly. 
Single-state simulations indeed show that the energy difference approaches zero for an increasing number of single-particle functions. These problems do not affect the validity of rescaling which is shown to work adequately and to be independent of the shape of the external potential. The energy differences compared to the analytical results range around about $1\%$.

For $h=10E_{0,H}$ a near degeneracy between the third and fourth state emerges. In contrast to the previously discussed degeneracy at vanishing barrier, it is very well described by the simulation. For large enough barrier height the double well resembles two separated anharmonic wells. According to the Bose-Fermi mapping, we identify the degenerate states by the superpositions of $(1,1,1, 0,\ldots); (1,0,1, 0,\ldots)$ and $(1,0,1, 0,\ldots); (1,1,1, 0,\ldots)$, showing the fermionic configurations in the left- and right-hand wells, respectively. Since these two states involve similar fermionic modes, their degeneracy is well described by the truncated basis-set expansion with a limited number of single-particle functions.

Overall, we find good description of the excited states by the rescaled simulation in the double well for a large variation of barrier height $h$ from a harmonic trap ($h=0$) to almost completely separated wells.

\section{\label{conclusion}Conclusion}

We have proposed a simple rescaling procedure for many-body calculations with a contact interaction in one dimension in a truncated Hilbert space
based on Eqs. (\ref{eq:rescaling}) and (\ref{eq:E0}). The method requires knowledge of only a single parameter. This is found from the interaction strength where the numerical ground-state energy matches the TG energy, which is easily found from single-particle calculations.
For two particles in a ring geometry the rescaling exactly reproduces the correct energies and expansion coefficients of the wave functions. Empirical evidence suggests that the scheme can be extended to  multiple particles, external trapping potentials, and low-lying excited states. Testing the method against exact results for ring geometries and harmonic and double-well potentials shows significant improvements in accuracy.

Because rescaling works for excited-state calculations and for arbitrary potentials it is expected to significantly improve time-dependent simulations as well, as long as the state of the system is well described by the low-lying part of the energy spectrum.

\section*{Acknowledgements}
We thank Peter Schmelcher  for discussions and encouragement and Sascha Z\"ollner for pointing out connections to $T$ matrix renormalization. T.E., D.W.H., J.G., and J.B. are supported by the Marsden Fund Council (Contract No. MAU0706) from government funding, administered by the
Royal Society of New Zealand. H.D.M. thanks the \emph{Deutsche Forschungsgemeinschaft} (DFG) for financial support.


%

\end{document}